\documentclass[]{aastex63}

\graphicspath{{./}{figures/}}
\usepackage{gensymb}
\accepted{May 3rd 2021}

\submitjournal{RNAAS}
\begin{document}
\title{Asteroid lightcurves and
detection, shape, and size biases
in large-scale surveys}

\correspondingauthor{Samuel Navarro-Meza}
\email{sn528@nau.edu}

\author[0000-0002-0289-4867]{Samuel Navarro-Meza}
\affiliation{Department of Astronomy and Planetary Science, Northern Arizona University, Flagstaff, AZ, 86011-6010, USA}
\affil{Instituto de Astronom\'{\i}a, Universidad Nacional Aut\'onoma de M\'exico, Ensenada B.C. 22860, M\'exico.}

\author[0000-0002-2894-6418]{Erin Aadland}
\affiliation{Department of Astronomy and Planetary Science, Northern Arizona University, Flagstaff, AZ, 86011-6010, USA}
\affiliation{Lowell Observatory, 1400 W Mars Hill Road, Flagstaff, AZ 86001, USA}

\author[0000-0003-4580-3790]{David Trilling}
\affiliation{Department of Astronomy and Planetary Science, Northern Arizona University, Flagstaff, AZ, 86011-6010, USA}

\begin{abstract}
    Most asteroids are somewhat elongated and have non-zero lightcurve amplitudes. Such asteroids can be detected in large-scale sky surveys even if their mean magnitudes are fainter than the stated sensitivity limits. We explore the detection of elongated asteroids under a set of idealized but useful approximations. We find that objects up to 1 magnitude fainter than a survey's sensitivity limit are likely to be detected, and that the effect is most pronounced for asteroids with lightcurve amplitudes 0.1--0.4~mag.
    This imposes a bias
    on the derived size and shape distributions of the population that must be properly accounted for.
\end{abstract}

\keywords{asteroids --- LSST --- surveys}

\section{Introduction}
The Vera C. Rubin Observatory will carry out the Legacy Survey of Space and Time (LSST), a 10 year survey that will produce hundreds of observations of both sidereal and non-sidereal sources.
One of the scientific goals of LSST is to inventory the Solar System by creating a comprehensive survey of small bodies. The LSST asteroid catalog will be the largest component of this database of moving objects, with over 5~million asteroids expected \citep{LSSTScienceBook}. 

The standard LSST observing cadence consists of 30~second exposures for a given field. To enable asteroid detection and establish proper motion vectors for the asteroids, each field will be observed twice per night.

In order for an asteroid to be detected, it has to have three observed pairs (a pair is defined as two observations on a given night) during a single lunation.

The elongation, or axial ratio, of an asteroid determines the amplitude of its lightcurve, and
 \cite{Erasmus2018}, and \cite{Mommert2018} found that main belt asteroids have average elongations in the range 0.74--0.80 (asteroids are typically modeled as ellipsoids with principal axes $a>b\geq c$, elongation being $b/a$). A non-spherical object --- most asteroids --- with an average magnitude near the limiting magnitude of the survey may be detected, or not, depending on the lightcurve phase at the time of observation.

\cite{Veres2017-biases} studied a variety of effects that could affect LSST's performance
in detecting Near Earth Objects. They made an implicit consideration of the asteroid's amplitude, and by using a fading function, they determined that NEOs will be detected by up to $\sim$0.5~magnitudes beyond the telescope's limiting magnitude. 

Here we make a simple quantitative study on the probability  of detecting elongated asteroids whose mean magnitudes are fainter than the limiting magnitude. Then we discuss on the implications of detecting aspherical asteroids for the scientific interpretation of the survey's catalogue. Our study is motivated by the upcoming LSST, but our results apply to any survey.

\section{Model Asteroid Lightcurves }

We modeled asteroid lightcurves using sine curves, allowing us to represent a set of observations in a simple yet realistic way.  The lightcurve amplitude is measured peak to trough, which makes it twice the mathematical amplitude of the sine curve. The sine's mean value represents the object's mean magnitude, which here is measured relative to the LSST's limiting magnitude.

We ignore asteroid rotation periods as the correspondence between observing time and rotational phase is assumed to be random and therefore is addressed in our random sampling (below).
We modeled 256 sine waves using 16~different relative magnitudes and 16~different amplitudes. The relative magnitudes are evenly distributed from 0 to 1.5. The amplitudes are evenly distributed in the range 0.05 to 1.55. 

To emulate an observing record, we take a random sample of 100 points of each sine wave, which we refer as ``observations.'' Each observation is assigned a random error that is drawn from a uniform distribution in the range [0.006,~0.1]~mag, to simulate the expected photometric error of a point source observed with the Vera Rubin Observatory, which will be dominated by a random error  \citep[see][]{LSSTScienceBook}. If 10\% of our observations are brighter than the limiting magnitude (relative magnitude of 0.0 mag), then we consider the asteroid as detected. 

To determine the effect the elongation of an object has on detection in a realistic scenario, we use main belt asteroids from the Lightcurve Database (LCDB\footnote{http://www.minorplanet.info/lightcurvedatabase.html}) \citep{Warner09}
to determine a probability function for lightcurve amplitudes. The “U code” is the LCDB's quality code to indicate the reliability of the lightcurve's period solution. It consists of 8~possible ratings between 0 (worst) and~3 (best). We take the reported maximum amplitude for cases with a U code of 3, 3-, and 2+. These are objects with a reliable period estimate, which is a proxy for the general quality of the lightcurve and the reliability of its amplitude.

At present, there are 2426~objects that meet these criteria, and we use the distribution of their amplitudes to assign the likelihood of finding an object of a given amplitude. However, these LCDB measurements are obtained at a range of phase angles, and the observed amplitude of an asteroid's lightcurve is dependent on the observation geometry such that larger phase angles exaggerate the amplitude: 
$\Delta m (0\degree) = \Delta m(\alpha)/(1 + s\alpha)$,
where $\Delta m$ is the amplitude, $\alpha$ the phase angle, and $s=0.019$ \citep[$s$ depends on composition; $s=0.019$ is an average value;][]{Zappala}. Main belt asteroids are observed at $\alpha \ < 30\degree$, so the correction factor is between 1 and 1.57; here we use an average correction factor of~1.2.

\begin{figure}
\begin{center}
\includegraphics[scale=0.6,angle=0]{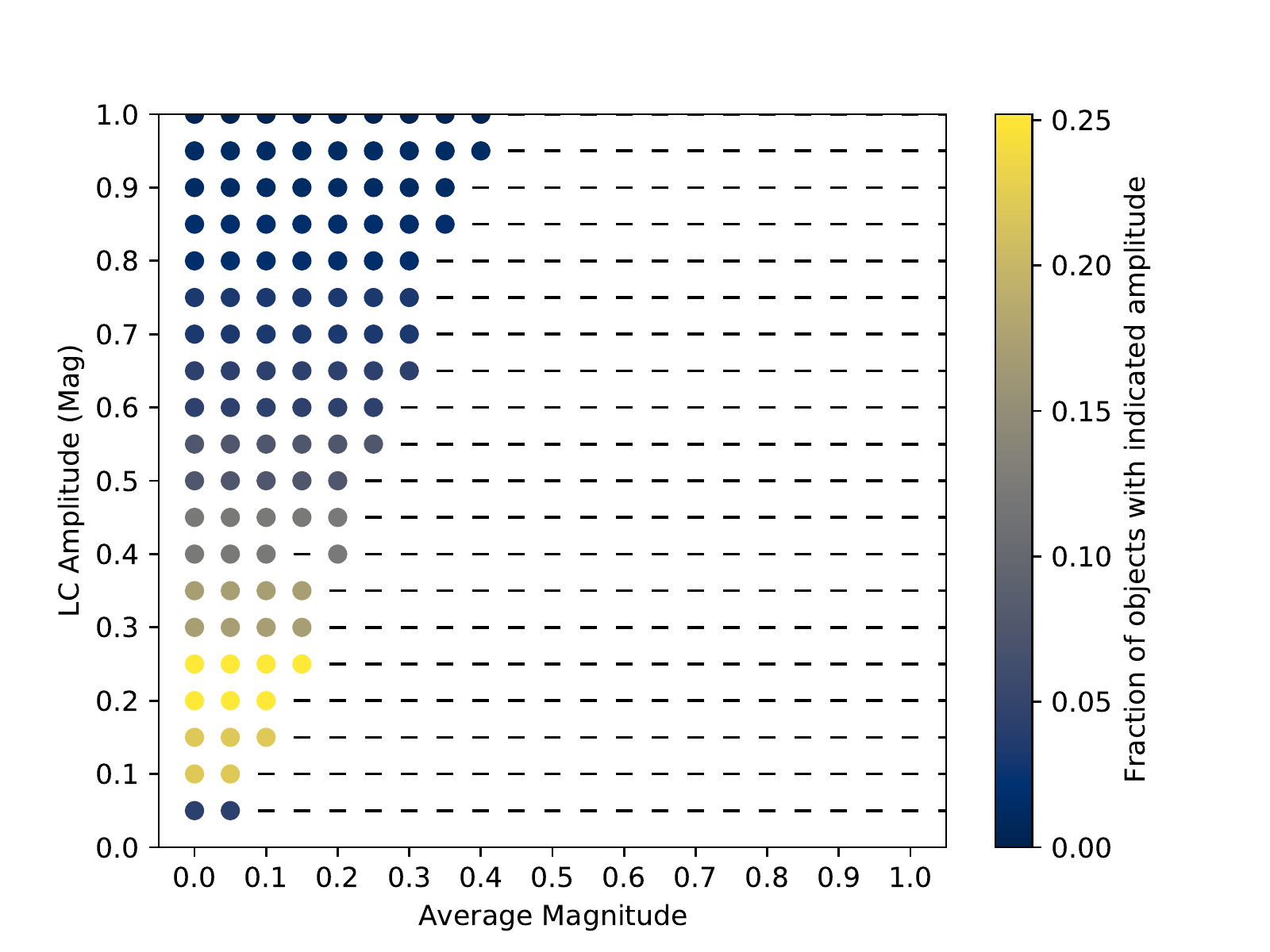}
\caption{Elongated asteroids detected in our simulation. Circles show combinations of relative magnitude and lightcurve amplitude that were detected in our model. Dashes represent cases modeled but not detected. 
Objects with amplitudes $>$1~magnitude were modeled and detected (as suggested by an extrapolation of the results shown here), but they are not shown here due to their low occurrence in the observed population (represented by the color scale).
\label{fig:dots}}
\end{center}
\end{figure}

Figure \ref{fig:dots} shows the summary of our results. The magnitude of an object that can be detected, relative to the nominal survey sensitivity, is approximately equal to $A/2$, where $A$ is the lightcurve amplitude. The color scale shows the fraction of real asteroids with a given amplitude according to the LCDB.

\section{Discussion} 
Elongated asteroids whose mean magnitudes are fainter than a survey's nominal sensitivity limit can still be detected. This effect is greatest for asteroids with large lightcurve amplitudes (highly elongated), but these are relatively rare in the known asteroid population. However, modest lightcurve amplitudes of 0.1--0.4~mag are common, and these asteroids will be frequently detected by LSST.

Therefore, the expected yield of LSST is probably larger than presently predicted, because most asteroids will have small/moderate lightcurve amplitudes. Furthermore, if a body has sufficient amplitude to become bright enough to be detected by LSST at its maximum brightness, only a small portion of the lightcurve will be observed. This will likely lead to the amplitude of the object being underestimated, and hence the mean (absolute) magnitude of the object also being underestimated. Any derived period may also be suspect.
Additionally, to determine the shape of an individual object, a reliable lightcurve is needed, and to estimate the elongation distribution for a population requires accurate partial lightcurves  \citep[see][and references therein]{McNeill2019}. During LSST's ten year survey many asteroids' sub-observer longitudes will change. This may provide the opportunity to disentangle lightcurve effects near the sensitivity limit, but could also result in asteroids that are detected in one epoch becoming undetectable in a different epoch.

Critically, therefore, we emphasize that detections of these elongated bodies can bias calculations of asteroid size distribution or shape distribution 
derived from the LSST asteroid catalog
if not properly accounted for.

\section{Conclusions}

We find that elongated asteroids whose mean magnitudes are fainter than a survey's detection limit can still be detected. This effect is most likely to be significant for asteroids with small or moderate lightcurve amplitudes (0.1--0.4~mag). However, this is enough to create a systematic bias on the
size and shape distributions derived from LSST, or indeed any large-scale survey.

\bibliographystyle{aasjournal}
\bibliography{main_aasformat}
\end{document}